\documentclass[final,5p,times,number]{elsarticle}

\usepackage{amssymb}
\usepackage{lipsum}
\usepackage{bm}

\makeatletter
\def\ps@pprintTitle{%
   \let\@oddhead\@empty
   \let\@evenhead\@empty
   \let\@oddfoot\@empty
   \let\@evenfoot\@empty
}
\makeatother

\begin{document}

\begin{frontmatter}

\title{Why Emergence of Gravity\\ in Matrix Theories is Entropic}


\author{Vatche Sahakian\fnref{footnote_label}}
\fntext[footnote_label]{Harvey Mudd College, Physics Department,  Claremont, CA 91711, \texttt{sahakian@hmc.edu}}

\begin{abstract}
Matrix theories exhibit the phenomenon of spacetime emergence in certain regimes of their dynamics. In this work, I posit that the key to this emergence is a hierarchy between two timescales -- very slow modes that an observer measures plus a chaotic cloud of very fast modes; this then leads to a natural operator algebra for measurements for the slow modes and an associated density matrix for the full system. I show that this density matrix implies the same gravitational potential energy that has been identified previously in the literature; furthermore, I show that the corresponding emergent gravitational force is nothing but the entropic force associated with the  entropy of the fast modes of the system. I also demonstrate that the posited regime corresponds to distances being super-Planckian in the dual eleven dimensional supergravity. 
\end{abstract}

%

\end{frontmatter}

\section{Introduction}
\label{sub:intro}

In a typical Matrix theory, the degrees of freedom of a quantum mechanics are packaged in $N\times N$ matrices and the theory has U(N) gauge symmetry~\cite{Banks:1996vh}~\cite{Berenstein:2002jq}. Consider a regime where any such matrix $\bm{X}$ can be arranged as follows
\begin{equation}
	\bm{X}\sim \left(
	\begin{array}{ccc}
	\fbox{\parbox[c][0.5cm][c]{0.5cm}{\centering  $\bm{D}_1$}} & {\bm{R}} \\
	{\bm{R}}^\dagger & \fbox{\parbox[c][1cm][c]{1cm}{\centering  $\bm{D}_2$}}
	\end{array}
	\right)
\end{equation}
where $\bm{D}_1$ and $\bm{D}_2$ describe block-diagonal slow modes, while the off-diagonal $\bm{R}$ describes fast modes. I will refer to the longer timescale associated with the diagonal blocks as $\tau_D$, and the fast timescale as $\tau_R$. We are then assuming that we are studying the theory in a regime where
\begin{equation}
	\tau_D \gg \tau_R\ .
\end{equation}
I will demonstrate that if this hierarchy of timescales is present, we are in a regime that is associated with emergent spacetime and gravity. We might worry about the U(N) transformations mixing up the separation of slow and fast modes and, correspondingly, that this statement is not consistent with the gauge symmetry. We will see that this is not the case when we discuss the operator algebra of physical measurements relevant to this regime.  

To be more specific, in this work I focus on the case where $N=2$; we will consider an SU(2) Matrix theory. While emergence of gravity is often studied in a regime where $N$ is large, it is also well-known that the correspondence between Matrix theories and gravity occurs at finite $N$, in the so-called DLCQ scaling regime~\cite{Bigatti:1997gm,Susskind:1997cw,Seiberg:1997ad}. I will comment about the implications to the large $N$ setting later. 

Finally, Matrix theories of common interest are often supersymmetric, and the supersymmetry plays a critical role in allowing for computations at weak coupling that can be extrapolated to strong coupling. While a matching between results from supergravity and Matrix theories necessitates the inclusion of the fermions in the Matrix theory
~\cite{Tafjord:1997bk}-\cite{Lee:2004kv}, 
the general scheme of how this matching happens can also be demonstrated only in the bosonic sector. Furthermore, as we will see, the presence of a hierarchy of timescales will lead us to propose a thermal density operator for the fast modes, a setting where supersymmetry would anyways be broken. In summary, I focus in this work on bosonic SU(2) Matrix theory, deferring a supersymmetric treatment and the large $N$ case to later. 

In the upcoming sections, I will demonstrate that the existence of a hierarchy between two timescales as described above results in a force law between slow degrees of freedom that is entropic in origin
\begin{equation}
	\bm{F} = T\,\bm{\nabla}S
\end{equation}
where $S$ is the entropy of the fast modes. I also show that this force is what has been identified in the literature as the emergent gravitational force from Matrix theories. Finally, we will see that the large separation between timescales is equivalent to the statement that the slow coordinates of the Matrices are separated by distances much greater than the eleven-dimensional Planck scale. We hence find out that emergence of gravity and spacetime in Matrix theories affirms the entropic nature of gravity, as advocated in~\cite{Verlinde:2010hp}.

\section{Bosonic Matrix Theory as the Playground}
\label{sub:bfss}

I consider the bosonic sector of the Banks-Fischler-Shenker-Susskind (BFSS) Matrix theory, described by the Lagrangian~\cite{Fuster:2005js}
\begin{equation}
	L=\frac{1}{2} {{\bm{{\dot{X}}}}^i_a}^2+\frac{g^2}{4} \bm{X}^i_a \bm{X}^j_b \bm{X}^i_a \bm{X}^j_b-\frac{g^2}{4} \bm{X}^i_a \bm{X}^j_b \bm{X}^i_b \bm{X}^j_a+i \bm{b}_a \dot{\bm{c}}_a\ .
\end{equation}
$i,j=1,\ldots, d$ is the target space index ($d=9$ for the full space), while $a,b=1,2,3$ are color indices of SU(2). I use the Pauli matrix basis with algebra
\begin{equation}
	\left[\tau^a, \tau^b\right] = 2\,i\,\varepsilon^{abc}\tau^c\ .
\end{equation}
$\bm{c}_a$ and $\bm{b}_a$ are {\em self-dual} ghost degrees of freedom, and we have 
\begin{equation}
	\left[\bm{X}^i_a, \bm{P}^j_b\right]=i \delta_{a b} \delta^{i j}, \quad\left\{\bm{c}_a, \bm{b}_b\right\}=\delta_{a b}\ .
\end{equation}
The BRST charge is given by~\cite{VanHolten:2001nj}
\begin{equation}
	\bm{\Omega}=\bm{c}^a \bm{G}^a-\frac{i g}{2} \epsilon^{a b c} \bm{c}^a \bm{c}^b \bm{b}^c .
\end{equation}
where the constraint is
\begin{equation}
	\bm{G}_a=g \epsilon_{a b c} {\bm{X}^i}_b \cdot {\bm{P}^i}_c\ ,
\end{equation}
corresponding to the static gauge
\begin{equation}
	A_0=0\ .
\end{equation}
In this form, the BRST auxiliary scalars have already been eliminated. The ground state of the system is defined by
\begin{equation}
	\bm{A}^i_a \Psi_0=\bm{b}^a \Psi_0=0
\end{equation}
where 
\begin{equation}
	\bm{A}_a^i=\frac{1}{\sqrt{2}}\left(\xi^{-1} \bm{X}_a^i+i \xi \bm{P}_a^i\right)\ .
\end{equation}
The role of the constant $\xi$ will become clear soon. To construct the physical states, first write 
\begin{equation}
	\Phi_0=\frac{1}{\sqrt{2}}\left(1+\frac{1}{3!} \bm{\theta}\, \epsilon^{a b c} \bm{c}^a \bm{c}^b \bm{c}^c\right) \Psi_0
\end{equation}
with $\bm{\theta}$ being yet another ghost variable;
then act on it with raising operators as needed
\begin{equation}
	\Phi[M]=M\left[\bm{A}^{\dagger}\right] \Phi_0 .
\end{equation}
where
\begin{equation}
	M\left[\bm{A}^{\dagger}\right]=\sum_n \mu_{a_1 \ldots a_n} \bm{A}_{a_1}^{\dagger} \ldots \bm{A}_{a_n}^{\dagger}
\end{equation}
with the $\mu_{a_1 \ldots a_n}$ being invariant SU(2) tensors. These can be constructed from products of traces of products of Pauli matrices
\begin{equation}
	\mbox{Tr} \left( \tau^{a_1} \tau^{a_2} \cdots \tau^{a_{i-1}}\right) \mbox{Tr} \left( \tau^{a_i} \tau^{a_{i+1}} \cdots \tau^{a_{j-1}}\right) \left( \tau^{a_j} \tau^{a_{j+1}} \cdots\right) \cdots
\end{equation}
All this requires that the measure in the Hilbert space is given by
\begin{equation}\label{eq:measure}
	\langle\Phi, \Psi\rangle = \int d \theta\, \prod_a d c^a \prod_{i, a} d X_i^a\, \Phi^{\dagger}[{X}, c]\, \Psi[{X}, c]\ .
\end{equation}
Note that $\bm{\theta}$ is not self-dual and comes in a pair with another ghost $\bm{\zeta}$, with $\{\bm{\theta}, \bm{\zeta}\}=1$.
Its role is to get the right fermion numbers for the physical states.

\section{Operator Algebra and Measurement}
\label{sub:operators}

The natural operator algebra that split our system's Hilbert space into the setup for slow diagonal modes versus fast off-diagonal ones can be written as~\cite{Mazenc:2019ety,Balachandran:2013cq,DaltonGoold}
\begin{equation}
	\mathcal{A} = \left\langle  | X \rangle_{D} \langle X'| \otimes \bm{1}_R\right\rangle
\end{equation}
where the subscript $D$ signifies the sector of the Hilbert space acted upon by operators in the direction of the third Pauli matrix
\begin{equation}\label{eq:tau3}
	\tau^3 = \left(\begin{array}{cc}1 & 0\\0 & -1\end{array}\right)\ \ \ ,\ \ \ \bm{\hat{X}}_3^i=\left(\begin{array}{cc}\bm{{X}}_3^i & 0\\0 & -\bm{X}_3^i\end{array}\right)\ ;
\end{equation}
and the $R$ subscript indicates the subspace of the off-diagonal modes in the directions of $\tau^1$ and $\tau^2$. But an SU(2) symmetry transformation mixes diagonal and off-diagonal modes, so the separation of the Hilbert implied by this algebra breaks the gauge symmetry. To remedy this, we proceed as follows.

The outcome of a measurement of the position and momentum of the slow degree of freedom is best represented by a coherent state~\cite{PhysRev.152.1103}. We then first construct the coherent state parameterized by the complex numbers $\alpha^i_a$ using the displacement operator $\bm{D}(\alpha)$ acting on the vacuum
\begin{eqnarray}
\bm{D}_\alpha | 0 \rangle &=&\exp \left\{\alpha^i_a{\bm{A}^i_a}^{\dagger}-\bar{\alpha}^i_a\, \bm{A}_a^i\right\} | 0 \rangle \nonumber \\ &\equiv& | \alpha \rangle\ .
\end{eqnarray}
This however is {\em not} a physical state. The action of SU(2) in the adjoint representation is isomorphic to the action of SO(3) in the fundamental. This means that we can make this state be physical if we add the following identification to the Hilbert space
\begin{equation}\label{eq:equivalence}
	|\alpha \rangle \approx | \beta \rangle\ \ \ \ \mbox{if ${\bar{\alpha}}_a^i \alpha_a^i = {{\bar{\beta}}}_a^i {\beta}_a^i$ (no sum over $i$, $\forall i$)}
\end{equation}
These rotations in color space mix the diagonal and off-diagonal modes, and the identification makes sure that we don't treat any direction in color space as special. Without loss of generality, we can then pick a representative of this equivalence class on the gauge orbit, say the one pointing in the $\tau^3$ direction. I write for notational purposes
\begin{equation}
	\bm{X}^i \equiv \bm{X}_3^i\ \ \ ,\ \ \ \bm{P}^i \equiv \bm{P}_{3}^i\ .
\end{equation}
We then have
\begin{equation}
	\bm{A}=\frac{1}{\sqrt{2}}\left(\xi^{-1} \bm{X}+i \xi \bm{P}\right)
\end{equation}
with now the constant $\xi$ defined as\footnote{A coherent state is one with $\Delta X \Delta P=\mbox{ minimum }$ and maximal entropy while the first and second moments of position and momentum are fixed. For an excellent review of this subject, see~\cite{nickwheeler}.}
\begin{equation}
	\xi\equiv\sqrt{\Delta X / \Delta P}
\end{equation}
where $\Delta X$ and $\Delta P$ relate to the resolution of our measuring instrument in position and momentum. I then use complex parameters $\alpha^i\equiv\alpha^i_3$ and write the corresponding state in the Hilbert space simply as
\begin{equation}\label{eq:displacement}
\bm{D}(\alpha) | 0 \rangle =\exp \left\{\alpha^i\, {{\bm{A}}^i}^{\dagger}-{\bar{\alpha}}^i\, {\bm{A}}^i\right\} | 0 \rangle = | \alpha \rangle
\end{equation}
The wavefunction is then given by 
\begin{eqnarray}
	\psi_\alpha(x)&=&\langle x| \alpha\rangle
	=\exp \left\{-\frac{i}{2} X_0^iP_0^i\right\} \cdot\left(\frac{1}{2 \pi(\Delta X)^2}\right)^{\frac{1}{4}} \nonumber \\ 
	&\times& \exp \left\{-\left(\frac{x^i-X_0^i}{2 \Delta X}\right)^2+i\, P_0^i x^i \right\}
\end{eqnarray}
with
\begin{equation}
	\alpha^i=\frac{1}{\sqrt{2}}\left(\xi^{-1} X_0^i+i\, \xi\, P_0^i\right)\ .
\end{equation}
So, this is a Gaussian at position $X_0^i$ and thrown with momentum $P_0^i$. Note that the real part of $\alpha$ is position and the imaginary part is momentum. It is also worthwhile noting that coherent states form an over-complete basis with
\begin{equation}
	\langle\beta | \alpha\rangle =e^{-\frac{1}{2}|\beta|^2-\frac{1}{2}|\alpha|^2+\bar{\beta} \alpha}
\end{equation}
and
\begin{eqnarray}
\langle\beta | \alpha\rangle  \approx & 0 \quad \mbox{ only when }|\beta-\alpha| \mbox{ is large }\ .
\end{eqnarray}
Yet we still have
\begin{equation}\label{eq:completeness}
	\left.\int | \alpha \rangle \frac{d^2 \alpha}{\pi}\langle  \alpha | =\bm{1} \quad , \quad d^2 \alpha=d \alpha_r d \alpha_i\right.
\end{equation}
Finally, note two useful properties satisfied by the displacement operator~\cite{nickwheeler}
\begin{equation}\label{eq:Didentities}
	\bm{D}_\alpha \bm{D}_\beta=e^{\frac{1}{2}(\alpha^i \bar{\beta}^i-\bar{\alpha}^i \beta^i)} \cdot \bm{D}_{\alpha+\beta}\ \ \ ,\ \ \ \bm{D}^{-1}_\alpha=\bm{D}_{-\alpha}\ .
\end{equation}

Coming back to the task at hand, we imagine an observer making a measurement of the position and momentum of the proposed slow mode, picking the $\tau^3$ direction as a representative, ending up with the density matrix
\begin{equation}\label{eq:alpha}
	\bm{\rho}_D = |\alpha\rangle\langle\alpha| \approx |\beta\rangle\langle\beta|
\end{equation}
with $\bar{\beta}_a^i\,\beta_a^i = \bar{\alpha}^i\alpha^i$ for all fixed $i$. The Hilbert space identification assures that this is a physical state. 

For notational simplicity, I will be dropping the dressing of the state with the BRST ghosts -- assured that the physical density matrix commutes with the BRST charge; implicitly, expectation values are computed with the proper measure~(\ref{eq:measure}) when the ghosts are included. Note that the state~(\ref{eq:alpha}) corresponds to two D0 branes separated by a distance $2\,X_0$ and relative momenta $2\,P_0$. These are to describe the slow dynamics over timescales I call $\tau_D$.

We then write the Hamiltonian of the system as
\begin{eqnarray}\label{eq:Hbefore}
	\bm{H} &=& \frac{{\bm{P}^i}^2}{2}  \nonumber\\
	&+& \frac{{\bm{P}_{A}^i}^2}{2}+\frac{1}{2} g^2
   {\bm{X}^i}^2 {\bm{X}_A^j}^2 -\frac{1}{2} g^2 \bm{X}_{i}
   \bm{X}_{j} {\bm{X}_A^i} {\bm{X}_A^j} \nonumber\\
   &+& \frac{1}{4} g^2 {\bm{X}_A^i}^2 {\bm{X}_B^j}^2 -\frac{1}{4} g^2 {\bm{X}_A^i} {\bm{X}_A^j} {\bm{X}_B^i} {\bm{X}_B^j} \nonumber \\
   &\equiv& \bm{H}_D+\bm{H}_R\ ,
\end{eqnarray}
where
\begin{equation}
	A, B, \ldots = 1,2\ ;
\end{equation}
and I wrote the kinetic energy of the slow modes as
\begin{equation}
	\bm{H}_D \equiv \frac{{\bm{P}^i}^2}{2}\ .
\end{equation}
Assuming that the $\bm{X}_A^i$'s are evolving at a much faster pace, at a timescale $\tau_R\ll \tau_D$, I now posit that the effective Hamiltonian for the system between measurements of the slow modes is given by 
\begin{eqnarray}\label{eq:Halpha}
	\bm{H}_{\alpha} &=& \frac{{\bm{P}^i}^2}{2} + \frac{{\bm{P}_{A}^i}^2}{2} \nonumber\\
	&+&\frac{1}{2} g^2\langle\alpha |
   {\bm{X}^i}^2 | \alpha \rangle {\bm{X}_A^j}^2 -\frac{1}{2} g^2\langle \alpha | \bm{X}_{i}
   \bm{X}_{j}| \alpha \rangle {\bm{X}_A^i} {\bm{X}_A^j} \nonumber\\
   &+& \frac{1}{4} g^2 {\bm{X}_A^i}^2 {\bm{X}_B^j}^2 -\frac{1}{4} g^2 {\bm{X}_A^i} {\bm{X}_A^j} {\bm{X}_B^i} {\bm{X}_B^j}
\end{eqnarray}
The key here is that the slow modes come into the fast dynamics adiabatically, so that we take the expectation value of the interaction terms in the coherent state $|\alpha\rangle$. $\bm{H}_R^\alpha$ is then defined as
\begin{equation}
	\bm{H}_\alpha \equiv \bm{H}_D+\bm{H}^\alpha_R\ .
\end{equation}
Under the assumption of the existence of a hierarchy between the two timescales $\tau_R$ and $\tau_D$, I propose that the full density matrix relevant to measurements of the slow modes takes the form 
\begin{equation}
	\bm{\rho} = | \alpha \rangle \langle \alpha | \otimes \bm{R}_\alpha
\end{equation}
where
\begin{equation}
	\bm{R}_\alpha = \frac{\bm{Z}_\alpha^T}{\mbox{Tr}_R \bm{Z}_\alpha^T}
\end{equation}
with
\begin{equation}
		\bm{Z}_\alpha^T = \exp{\left\{-\frac{1}{T} \bm{H}_R^\alpha \right\}}
\end{equation}
living in the Hilbert subspace of the fast modes.
We are essentially saying that the fast modes are thermalized by chaotic dynamics at some temperature $T$. So, our system divides into two sectors: one with slow dynamics that acts as a reservoir to the other, providing it a temperature $T$ and external parameter $\alpha$; and another scrambled thermal sector for which the evolution of the slow modes is adiabatic. There is however a subtle point in this otherwise natural proposition: for a typical $N\times N$ matrix, the fast modes would be of order $N^2$ in number, while the slow modes would be of order $N$. So, it might seem that we are treating the smaller system as the larger one... This is not the case because the slow modes will typically access an infinite phase space, with $X_0$ and $P_0$ being unbounded. On the other hand, the fast system will have its degrees of freedom confined in phase space because, as we will soon see, they will be associated with high frequency oscillator-like excitations.   

Finally, let us note that the proposed density matrix is normalized $\mbox{Tr}\, \bm{\rho} = 1$ and mixed $\mbox{Tr}\, \bm{\rho}^2 < 1$. And if we were to trace over the fast modes, we get $\bm{\rho}_D = \mbox{Tr}_R\, \bm{\rho} = | \alpha \rangle \langle  \alpha |$ which is a mixed state due to the over-completeness of the coherent states.

\section{The Two Timescales}
\label{sub:twotimescales}

Let us take a step back and analyze the physical meaning of the hierarchy between the proposed two timescales. First, note that in our notation, the units of length for the variables I use are 
\begin{equation}
	[\bm{X}]\sim \ell^{1/2}\ \ \ ,\ \ \ [g]^2\sim \ell^{-3}
\end{equation}
The slow timescale is set by  
\begin{equation}
	\tau_D \simeq X_0^2\ .
\end{equation}
The fast timescale can be read off the Hamiltonian~(\ref{eq:Halpha}) from the second line:
\begin{equation}\label{eq:tauR}
	\tau_R \simeq \frac{1}{g\,X_0}
\end{equation}
which appears as an oscillator frequency, confining the phase space of the fast modes. We then need  $\tau_D\gg\tau_R$, implying
\begin{equation}\label{eq:regime}
	\tau_R^3 \ll \frac{1}{g^2}\ \ \ ,\ \ \ g\,X_0^3\gg1\ .
\end{equation}

Now, the conventions used in writing the Hamiltonian~(\ref{eq:Hbefore}) can be related to the IIA string theory and M-theory parameters through a sequence of dualities that relate DCLQ (or large $N$) Matrix theory to light-cone M-theory. One easily gets~\cite{Martinec:1998ja,Sahakian:2000bg,polchinski}
\begin{equation}
	X_0 =  \frac{r}{\sqrt{g_s \ell_s}}\ \ \ ,\ \ \ g^2 = \frac{g_s}{4\pi^2 \ell_s^3}
\end{equation}
where $r$ is now the physical separation between our two D0 branes in eleven dimensional M-theory units. We then have the statement $\tau_R\ll\tau_D$ transform into
\begin{equation}
	r \gg \ell_P
\end{equation}
where $\ell_P = g_s^{1/3}\ell_s$ is the eleven-dimensional M-theory Planck scale. So, the proposed hierarchy of timescales translates to saying that the two slow D0 branes that Matrix theory is describing are separated by a distance much larger than the Planck scale. This is indeed the regime we expect to identify with emergence of spacetime. 

\section{Matrix Entropic Force}
\label{sub:entropic}

To complete the picture and demonstrate that our treatment is consistent with the literature, I want to compute the force experienced by the slow degrees of freedom due to the cloud of fast off-diagonal modes. In general, we have
\begin{equation}
	\bm{\dot{\rho}}=-i[\bm{H}, \bm{\rho}]
\end{equation}
I first employ a basic technique commonly used in quantum optics~\cite{HowardCarmichael}: we switch to a Heisenberg picture with respect to $\bm{H}_D$
\begin{equation}
	{\bm{\tilde{\rho}}}=e^{i\,\bm{H}_D t}\, \bm{\rho}\, e^{-i\, \bm{H}_D t}\ .
\end{equation}
This trick allows us to zero onto the $\bm{H}_R$ from~(\ref{eq:Hbefore})
\begin{equation}
	\frac{d{\bm{\tilde{\rho}}}}{dt}=-i[\bm{\tilde{H}_R}, \bm{\tilde{\rho}}]\ .
\end{equation}
To compute the mutual force on the two D0 branes, we write
\begin{eqnarray}
	\frac{d\langle \bm{P}^i\rangle }{dt} &=& \frac{d}{d t} \mbox{Tr}(\bm{P}^i\,\bm{\rho})=\mbox{Tr}\left(\bm{\tilde{P}}^i \frac{d \bm{\tilde{\rho}}}{d t}\right) \nonumber \\ &=& -i\,\mbox{Tr}_S \bm{\tilde{P}}^i \mbox{Tr}_R \left[\bm{\tilde{H}}_R, \bm{\tilde{\rho}}\right]
\end{eqnarray}
Through a series of manipulations involving inserting the identity strategically in various places, we get
\begin{eqnarray}
	\frac{d\langle \bm{P}^i\rangle }{dt} &=& -i\,\int d^d x\,d^2\beta\, \langle \alpha | x \rangle \langle x | \bm{P}^i | \beta \rangle \langle \beta | \bm{H}_R \bm{R}_\alpha | \alpha \rangle \\
		&+& i\,\int d^d x\,d^2\beta\, \langle \beta | x \rangle \langle x | \bm{P}^i | \alpha \rangle \langle \alpha | \bm{H}_R \bm{R}_\alpha | \beta \rangle \\
		&=&-i\, \mbox{Tr}_R \langle \alpha |\left[\bm{P}^i, \bm{H}_R\right] | \alpha \rangle \bm{R}_\alpha \\ 
		&=& - \mbox{Tr}_R \bm{\nabla_X}\bm{H}_R^\alpha\bm{R}_\alpha
\end{eqnarray}
where in the last step, I made use of the fact that $\bm{H}_D\propto \bm{P}^2$. Note that the $e^{\pm i\,\bm{H}_D t}$ also cancel\footnote{Note also that this technique can be employed to build up an analytical perturbation series in $\bm{H}_R$; as we will be interested in strongly coupled dynamics, I do not develop this direction here.}. The derivative in the last line is defined as
\begin{equation}
	\bm{\nabla_X}\bm{H}_R^\alpha \equiv \lim_{\Delta X_0\rightarrow 0} \frac{\bm{H}_R^{\alpha+\xi^{-1}\Delta X_0/\sqrt{2}} - \bm{H}_R^\alpha}{\Delta X_0}
\end{equation}

This is a very satisfying result. In the limit where the temperature goes to zero, $T\rightarrow 0$, the trace involving $\bm{R}_\alpha$ is dominated by the ground state of $\bm{H}_\alpha$
\begin{equation}
	- \lim_{T\rightarrow 0}\mbox{Tr}_R \bm{\nabla_X}\bm{H}_R^\alpha\bm{R}_\alpha \rightarrow -\frac{d E_0^\alpha}{dX_0}
\end{equation}
This means that the force is the gradient of the ground state energy of the effective Hamiltonian $\bm{H}_\alpha$. This is precisely the conclusion of computing the gravitation force in Matrix theories using other methods (background field gauge, one-loop path integral evaluation of the effective potential)
~\cite{Tafjord:1997bk}-\cite{Lee:2004kv}. 
So, the force law we computed is generically what has been shown, in cases where supersymmetry and associated non-renormalization theorems help keep the computation under control, to be the gravitational force in eleven dimensional light-cone supergravity. 

Our next goal is to relate this to the entropy of the fast modes. To do this, I first note that the energy of the system is given by
\begin{equation}
	E(X_0) = \mbox{Tr}\,\bm{\rho} \bm{H}_D + \mbox{Tr}\,\bm{\rho} \bm{H}_R^\alpha
\end{equation}
By energy conservation, we have\footnote{The energy is a function of $X_0$ and does not depend on $P_0$. In Matrix theories with supersymmetry, the contribution from the fermions typically adds momentum dependence as well. We would correspondingly need an expanded version of this equation.}
\begin{equation}
	0 = \frac{d E(X_0)}{dt} = \frac{d E(X_0)}{dX_0^i} \frac{dX_0^i}{dt}
\end{equation}
This implies that a change due to a shift in the $X_0$ coordinate (the relative separation between the slow objects) is given by
\begin{equation}
	\bm{\nabla_X} E = 0 = \mbox{Tr}\,(\bm{\nabla_X}\bm{\rho})\bm{H}_D + \mbox{Tr}\,(\bm{\nabla_X}\bm{\rho}) \bm{H}_R^\alpha + \mbox{Tr}\,\bm{\rho}\, \bm{\nabla_X}\bm{H}_R^\alpha
\end{equation}
The first term vanishes because
\begin{equation}
	\mbox{Tr}_D |\alpha+\Delta X_0 \rangle \langle \alpha+\Delta X_0 |\bm{H}_D-\mbox{Tr}_D |\alpha \rangle \langle \alpha |\bm{H}_D = 0
\end{equation}
remembering that $\bm{H}_D\propto \bm{P}^2$. We then have
\begin{equation}\label{eq:iden}
	\mbox{Tr}_R (\bm{\nabla_X} \bm{R}_\alpha) \bm{H}_R^\alpha+\mbox{Tr}_R \bm{R}_\alpha (\bm{\nabla_X} \bm{H}_R^\alpha) = 0\ .
\end{equation}

Now, let us look at the Von Neumann entropy of our density matrix. We have
\begin{equation}
	S=-\mbox{Tr}\, \bm{\rho} \ln \bm{\rho}
\end{equation}
which gives
\begin{equation}
	S = -\langle \alpha |\left(\ln |\alpha\rangle\langle \alpha|\right)|\alpha\rangle - \mbox{Tr}_R \bm{R}_\alpha \ln \bm{R}_\alpha = S_D+S_R
\end{equation}
We want however the {\em change} in the entropy; that is generically
\begin{equation}
	\Delta S=-\mbox{Tr} \Delta \bm{\rho} \ln \bm{\rho} - \mbox{Tr} \Delta \bm{\rho} = - \mbox{Tr} \Delta \bm{\rho} \ln \bm{\rho}
\end{equation}
where we used $\mbox{Tr}\bm{\rho} = 1$. So, we have
\begin{equation}
	\Delta S = \Delta S_D+\Delta S_R
\end{equation}
Now, the first piece can be computed as follows, using the displacement operator from before (equation~(\ref{eq:displacement}))
\begin{eqnarray}
	\Delta S_D &=& -\mbox{Tr}\, \bm{D}_{\Delta \alpha}\bm{D}_\alpha|0\rangle\langle 0 | \bm{D}_{-\alpha}\bm{D}_{-\Delta \alpha} (\ln |\alpha\rangle\langle \alpha |) - S_D \nonumber \\
	&=& - (\alpha \Delta \bar{\alpha} + \bar{\alpha}\Delta\alpha)\, S_D  \nonumber \\ 
	&-& \Delta \alpha\,\mbox{Tr}_D\, \bm{A}^\dagger |\alpha\rangle\langle \alpha |\,(\ln |\alpha\rangle\langle \alpha |) \nonumber \\
	&-&\Delta \bar{\alpha}\,\mbox{Tr}_D\,  |\alpha\rangle\langle \alpha |\bm{A}\,(\ln |\alpha\rangle\langle \alpha |)\ ,
\end{eqnarray}
where I used~(\ref{eq:Didentities}) and the fact that $|\Delta\alpha|$ is assumed to be small.
One can then write
\begin{equation}
	\ln |\alpha\rangle\langle \alpha| = \ln \left(\bm{1}+(|\alpha\rangle\langle \alpha|-\bm{1})\right)
\end{equation}
We can now substitute the Taylor expansion of the logarithm since we are assured convergence due to~(\ref{eq:completeness}). We then easily get
\begin{equation}
	\Delta S_D = 0\ .
\end{equation}
We can understand this from the fact that the coherent state is known to be a state of maximal entropy~\cite{PhysRev.152.1103}. 
We then have the change in the total entropy given by the change in the entropy of the fast modes
\begin{equation}
	\Delta S=\Delta S_R = -\mbox{Tr}_R \Delta \bm{R}_\alpha \ln \bm{R}_\alpha\ .
\end{equation}
Using 
\begin{equation}
	\ln \bm{R}_\alpha = -\frac{1}{T}\bm{H}_R^\alpha - \ln \mbox{Tr}_R e^{-\bm{H}_R^\alpha/T}
\end{equation}
we then get
\begin{eqnarray}
\Delta S &=& \frac{1}{T}\mbox{Tr}_R \Delta \bm{R}_\alpha \bm{H}_R^\alpha \Rightarrow \bm{\nabla}_X S = \frac{1}{T}\mbox{Tr}_R (\bm{\nabla}_X \bm{R}_\alpha) \bm{H}_R^\alpha \nonumber \\
&=& - \frac{1}{T} \mbox{Tr}_R \bm{\nabla_X}\bm{H}_R^\alpha\bm{R}_\alpha
\end{eqnarray}
using~(\ref{eq:iden}).
Hence, we have shown
\begin{equation}
	\frac{d\langle \bm{P}^i\rangle}{dt}=+T
\,\bm{\nabla}_i S\ .
\end{equation}
The gravitational force law computed from Matrix theories is nothing but the entropic force due to thermal fast modes, in relation to the slow modes that the classical observer would measure. I also note that the motion of the slow modes leads to the increase in the entropy of the system.

\section{Conclusions and Outlook}\label{sec:conclusion}

The approach I have proposed in this work -- a certain grouping of the Hilbert space and an associated density matrix that is most relevant to the mechanism of emergence of space and gravity in Matrix theories -- differs from some of the recent propositions in the literature~\cite{Das:2020xoa,Das:2020jhy,Hampapura:2020hfg}. In these works, target space entanglement is adopted as the strategy for understanding how gravity is encoded in the matrices of Matrix theories. Our proposition is to instead focus on entanglement between sub-blocks of matrices instead, guided by a separation of two timescales in the dynamics, and framed in manner that respects the gauge symmetry and the physical space of states\footnote{This idea was inspired by previous computations~\cite{Sahakian:2019cxc}~\cite{Gray:2020zov}. I also note some similarities of our proposal with some of the suggestions in the work of~\cite{Gautam:2022akq}.}. I have successfully demonstrated that this approach leads to a simple and direct transcription of gravity and spacetime data into quantum entanglement of matrices, in tune with the broader proposition of entropic gravity by~\cite{Verlinde:2010hp} (see also~\cite{Jacobson:1995ab}).

Our system is governed by two dimensionless effective couplings, determined from $g$ and the two relevant timescales
\begin{equation}
	G_{eff}^2 \equiv g^2\tau_D^3 = \left(g\,X_0^3\right)^2\ \ \ ,\ \ \ g_{eff}^2 \equiv g^2\tau_R^3 = \left(g\,X_0^3\right)^{-1}\ .
\end{equation}
This implies that, in the regime $\tau_D\gg\tau_R$ as given by~(\ref{eq:regime}) -- $g\,X_0^3 \gg 1$, the dynamics of the slow modes is strongly coupled while that of the fast modes is weakly coupled -- yet highly non-linear and chaotic. All this suggests that a full treatment of the dynamics will require numerical methods as the system is otherwise too complex. Naturally, emergence occurs around $g\,X_0 \gtrsim 1\Rightarrow r \gtrsim \ell_P$, when the separation between the D0 branes becomes greater than the eleven dimensional Planck scale, which is also the onset of the regime of a hierarchy between two timescales in the Matrix dynamics. 
 
The system of fast thermal matrix degrees of freedom is associated with two macro-parameters, a temperature $T$ and the position/momentum of the slow objects, packaged in the coherent state complex parameter $\alpha$. The latter is like a volume parameter for the thermal system that is introduced into the dynamics by virtue of the act of measurement by an external agent -- the observer measuring the position and momentum of the slow degree of freedom. Yet, it is important to remember that the entire system of slow plus fast modes is closed, and we can also devise a micro-canonical approach -- which allows us to determine $T$ as a Lagrange multiplier. We would first need to compute the free energy of the fast modes
\begin{equation}
	F(T,\alpha) = - T\,\ln \mbox{Tr}_R e^{-\bm{H}_R^\alpha/T}\ .
\end{equation}
This can only be done numerically given the strong coupling at play in the system. 
We would then compute the entropy and invert it $S(T,\alpha) = -{\partial F}/{\partial T} \Rightarrow T(S,\alpha)$.
Finally, we apply the inverse Legendre transform, all numerically, and get the $
	E(S,\alpha) = F(T(S,\alpha))+T(S,\alpha)\, S$. This finally allows us to compute the temperature from
 $T(S,\alpha) = ({\partial E}/{\partial S})_\alpha$.
 
 I can provide a good guess of what to expect here. The fast timescale $\tau_R$ should be related to the fast scrambling timescale of Matrix theories, an inherent property of the internal chaotic dynamics
 \begin{equation}
 	\tau_R\sim \tau_{\tiny scrambling}\ .
 \end{equation}
 We know that the scrambling timescale is given by~\cite{Sekino:2008he}
\begin{equation}
	\frac{1}{T} \sim \frac{\tau_{\tiny scrambling}}{\ln S}
\end{equation}
Noting~(\ref{eq:tauR}), we then have
\begin{equation}
	T \simeq g\,X_0\,{\ln S} \simeq  g\,\xi\,\mbox{Re}(\alpha)\,{\ln S}\ ,
\end{equation}
where $\xi$ depends on the resolution of the measuring instrument we use to read off the position and momentum of the slow D0 branes\footnote{It is interesting to see the role of the measuring process in the emergence phenomenon. For example, an infinitely accurate position measurement, $\xi\rightarrow 0$, would lead to the zero temperature limit at constant entropy. Looking at the quadratic part of the effective Hamiltonian~(\ref{eq:Halpha}), the eigenfrequencies can be found to be $\omega_1 = g\,\sqrt{\xi\,(d-1)}/2$ for two modes ($d$ being the dimension of the target space), and $d-2$ modes with $\omega_2 = \sqrt{g^2\xi (d-1)/2+g^2X_0^2}\simeq g\,X_0$. $\omega_1 \rightarrow 0$ is typically identified with the zero modes arising from commuting matrices, a well-known instability of BFSS Matrix theory~\cite{Sahakian:2000bg,Sahakian:2001zs,Maldacena:2001ky}. We see that, using a more realistic measurement paradigm that introduces coherent states also naturally regularizes this instability.}. Note that $S$ is the entropy of the fast modes only. A more interesting way to write this relation would be
\begin{equation}
	T\,g^{-2/3}\simeq (g^{1/3} X_0) \ln S = G_{eff}^{1/3} \ln S
\end{equation}
since the energy scale is set by $g^{-2/3}$ (translating to $1/p_{11}=R_{11}/N$ in M-theory light-cone units for finite $N$) . This makes it transparent that, in the regime where all this is valid and we expect emergence, the coupling of interest is $G_{eff}=g\,X_0^3\gtrsim 1$. It would obviously be interesting to check these relations by computing the micro-canonical ensemble using numerical methods. We plan to report on this in a separate work~\cite{korinsahakian}, along with the inclusion of supersymmetry in the analysis. 

For $N>2$, or the large $N$ limit, I expect the general mechanics to remain the same, with factors of $N$ inserted wherever we have $g^2$. There is however an additional structure that will be present, which can be viewed as self-gravitational effects within sub-blocks of matrices describing two gravitating objects.  

Finally, I end with a few notes about the equivalence principle from the perspective of Matrix theory. Matrix theory is a {\em background dependent} formulation of quantum gravity, building up curved space from the flat Minkowski spacetime of light-cone M-theory. Understanding the underlying background independence can reveal the deep underpinnings of the emergence phenomenon. In the case at hand, where we compute the force experienced by slow D0 branes in the cloud of fast thermal matrix modes, an alternate approach might be to devise on the Matrix theory side the analogue of switching to Fermi normal coordinates -- observing the dynamics from the perspective of the freely falling frame of say one of the two gravitating objects~\cite{Gray:2020zov}. This will likely involve a canonical transformation in Matrix theory phase space -- noting that M-theory light-cone Galilean boosts are not enough since they lie in the U(1) of the gauge group. I hope to report on this as well in a separate work. 

\section*{Acknowledgments}

This work was supported by NSF grant number PHY-2109420 and the Burton Bettinger fund. I also thank Cameron Gray for very helpful discussions.

\section*{References}



\end{document}